\documentclass[a4,7pt]{article}
\usepackage[a4paper,hmargin={1.3cm,1.3cm},vmargin={1.5cm,1.5cm}]{geometry}
\usepackage{multicol} 

\usepackage[none]{hyphenat} 
\usepackage{amsmath}
\usepackage{amsfonts}
\usepackage{gensymb}

\usepackage{graphicx}
\usepackage{float}
\graphicspath{ {} }
\usepackage[labelfont=bf]{caption}

\usepackage[square, numbers, comma, sort&compress]{natbib} 

\usepackage{titling}
\posttitle{\par\end{center}}
\title{\textbf{Toward a Deterministic Nucleation Theory for Chirality-Controlled Nanotube Synthesis}}

\usepackage[affil-it]{authblk}

\author[]{Zhengrong Guo\thanks{zhengrong\_guo@yulinu.edu.cn}}

\affil[]{School of Energy Engineering, Yulin University,  Yulin, 719000, Shaanxi, People's Republic of China}

\begin{document}

\maketitle
\begin{abstract}
The electronic properties of carbon nanotubes are governed by their chirality, specified by the integer indices $(n,m)$. While chirality-controlled synthesis has achieved notable successes, theoretical understanding remains predominantly focused on post-nucleation growth. Two fundamental obstacles impede deeper insight: the absence of a clear description of nucleation cap topology and its connection to tube chirality, and an incomplete understanding of atomic-level mechanisms governing templated cap formation. Here we address these challenges directly. First, we develop a mathematically rigorous topological framework for carbon networks that provides both a concise definition of cap structures and a quantitative relationship between cap architecture and chirality—the vector sum rule. Second, contrary to conventional perspectives attributing chirality enrichment to edge matching during growth, we demonstrate that chirality is deterministically encoded during nucleation through selective formation of specific cap structures on catalyst surfaces. For the specific case of $(12,6)$ nanotubes, we show that their enrichment arises from a six-fold symmetric cap with epitaxial matching to catalyst facets. Our deterministic nucleation theory not only provides a coherent explanation for chirality enrichment but also elucidates its pattern in chirality space. This work establishes a theoretical framework that redefines the field, shifting the paradigm from stochastic growth kinetics to deterministic nucleation programming and paving the way toward predictable synthesis. 
\end{abstract}

\begin{multicols}{2}

Carbon nanotubes (CNTs) hold exceptional promise for semiconductor applications due to their extraordinary electrical conductivity and chirality-dependent band gaps~\cite{Iijima1991,Dresselhaus1995}. However, achieving chiral-selective synthesis remains a fundamental challenge in nanotechnology~\cite{Li2024,Yang2022,Hersam2009_Sorting}. Chemical vapor deposition with solid catalysts offers a scalable and promising approach toward chirality-specific growth, having produced several CNTs with remarkable enantiomeric purity—including $(12,6)$ at 92\% and $(14,4)$ at 97\%~\cite{Yang2014,Yang2016}. Despite these advances, the physical principles governing chiral preference remain poorly understood. Moreover, the absence of a predictive theory connecting catalysts to resulting chiralities has impeded systematic extension of this approach to other chiralities~\cite{Brinkmann1999,Li2004,Ding2004,Harutyunyan2009,Liu2012,Xu2018}.  Nanotube growth proceeds via crystalline elongation from a nucleation cap—a curved graphene dome formed by incorporating six pentagons into a hexagonal lattice~\cite{Thess1996}. This cap architecture encodes complete chirality information within its topological structure~\cite{Artyukhov2014,Liu2017}. Developing a predictive theory connecting solid catalysts to the chiral outcomes requires addressing two interconnected challenges: deciphering chirality from cap topology, and elucidating how catalysts direct cap formation during nucleation. The first is a problem of topological interpretation, while the second involves complex physics of nucleation and catalyst-carbon interactions. Resolving these intertwined problems is essential for advancing this important field.
    
Here, we address the first challenge by introducing a topological theory that furnishes a mathematically rigorous connection between cap architecture and chirality—expressed as the vector sum rule. Building on this theoretical foundation, we then examine the second challenge through the lens of existing experimental evidence, proposing that chiral-selective growth on solid catalysts may be guided by structural matching between the carbon cap and catalyst surface. This integrated approach offers a coherent mechanistic framework for understanding the preferential formation of experimentally observed chiralities such as $(12,6)$ and $(14,4)$.

\section*{Vector Sum Rule}

\begin{figure}[H]
\includegraphics[width=1.0\linewidth]{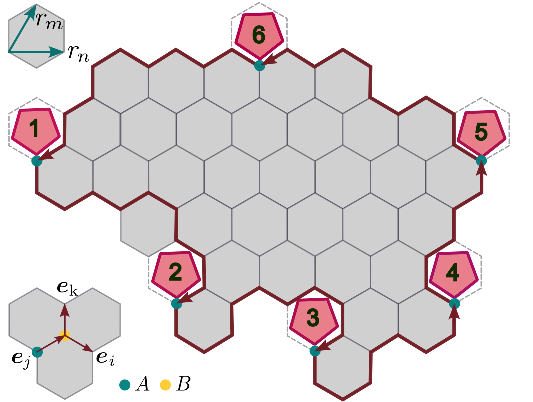}
\caption{Schematic of a carbon cap structure. Top left: the lattice basis vectors. Bottom left: the three topological basis vectors used in our framework.  Red pentagons mark the positions of the pentagon carbon rings.}
\label{fig:1}
\end{figure}

To establish a rigorous connection between cap structure and nanotube chirality, we develop a topological framework that captures the essential geometry of carbon networks. A carbon nanotube is fundamentally a two-dimensional surface curved in three-dimensional space~\cite{Andova2016}. Within this framework, we define three basis topological vectors $\mathbf{e}_{i}$, $\mathbf{e}_{j}$, $\mathbf{e}_{k}$ (Fig.1) and their respective inverses $\mathbf{e}_{i}^{-1}$, $\mathbf{e}_{j}^{-1}$, $\mathbf{e}_{k}^{-1}$. These vectors can be combined through a connection operation; for instance, $\mathbf{e}_{j} \cdot \mathbf{e}_{i}$ represents a basic building block, and any vector combined with its inverse yields the identity element $\mathbf{e}_{0}$. Repeated connections form topological paths, such as the closed path around a hexagonal ring:
\[
{\leftrightarrow}{\mathbf{P}}_{6} = \mathbf{e}_{j} \cdot \mathbf{e}_{k} \cdot \mathbf{e}_{i}^{-1} \cdot \mathbf{e}_{j}^{-1} \cdot \mathbf{e}_{k}^{-1} \cdot \mathbf{e}_{i}.
\]
	
The set $S$, comprising the identity, all basis vectors, their inverses, and finite combinations under connection, contains every allowed topological path in graphene-like networks. To incorporate sublattice-mismatching defects, we define left and right rotation operators   $\mathcal{L}$ and $\mathcal{R}$ that act on paths to introduce dislocations. These operators act cyclically on the basis vectors:	
\begin{align*}
	\mathcal{L}&: \mathbf{e}_{j} = \mathcal{L}(\mathbf{e}_{i}),\quad \mathbf{e}_{k} = \mathcal{L}(\mathbf{e}_{j}),\quad \mathbf{e}_{i}^{-1} = \mathcal{L}(\mathbf{e}_{k}) \\
	\mathcal{R}&: \mathbf{e}_{k}^{-1} = \mathcal{R}(\mathbf{e}_{i}),\quad \mathbf{e}_{i} = \mathcal{R}(\mathbf{e}_{j}),\quad \mathbf{e}_{j} = \mathcal{R}(\mathbf{e}_{k})
\end{align*}
	
Although $S$ itself does not form a group, the subset generated from $\mathbf{e}_{0}$, $\mathbf{p}_{ji} = \mathbf{e}_{i} \cdot \mathbf{e}_{j}$, and $\mathbf{p}_{jk} = \mathbf{e}_{j} \cdot \mathbf{e}_{k}$ constitutes an Abelian group $V = \{\mathbf{e}_{0}, \mathbf{p}_{ji}, \mathbf{p}_{jk}, \ldots\}^+$, wherein the connection operation is commutative and invertible. Within $V$, we adopt the plus sign to denote the connection operation between elements.
	
In graphene or nanotubes, any topological path connecting crystallographically equivalent atoms ($A\rightarrow A$ or $B\rightarrow B$) belongs to the group $V$. The value of such a path is defined as the element $\alpha \mathbf{p}_{ji} + \beta \mathbf{p}_{jk}$ (with $\alpha, \beta \in \mathbb{Z} $) to which it is homotopic. This leads to a purely topological definition of chirality: for a carbon nanotube of chirality $(n, m)$, a closed path encircling the circumference takes the value $n \mathbf{p}_{ji} + m \mathbf{p}_{jk}$ (see Supplemental Material for detailed description of the topological framework). 
	
We now incorporate pentagons to transform a closed graphene path into a nanotube path. Each pentagon $t$ ($t = 1, \ldots, 6$) occupies a lattice position $(n_t, m_t)$, represented topologically as $n_t \mathbf{p}_{ji} + m_t \mathbf{p}_{jk}$. The path from pentagon $t$ to pentagon $t+1$ (with cyclic indexing, $t+1 \equiv 1$ for $t=6$) is given by:	
\[
\mathbf{P}_{t\to t+1} = (n_{t+1} - n_t)\mathbf{p}_{ji} + (m_{t+1} - m_t)\mathbf{p}_{jk},
\]	
Together, these form a closed graphene path summing to $\mathbf{e}_{0}$:
\[
\overset{\leftrightarrow}{\mathbf{P}}_{\text{Graphene}} = \sum_{t=1}^{6} \mathbf{P}_{t \to t+1} = \mathbf{e}_0.
\]	
Introducing a pentagon---a topological kink represented by the path
\[
\mathbf{P}_5 = \mathbf{e}_{j} \cdot \mathbf{e}_{k} \cdot \mathbf{e}_{i}^{-1} \cdot \mathbf{e}_{j}^{-1} \cdot \mathbf{e}_{k}^{-1} = \mathbf{e}_{i}^{-1}
\]
---breaks this closure. Although $\mathbf{P}_{1 \to 2}$ connects directly to $\mathbf{P}_5$, the combined path $\mathbf{P}_{1 \to 2} + \mathbf{P}_5$ cannot connect directly to $\mathbf{P}_{2 \to 3}$ due to a fundamental sublattice mismatch introduced by the pentagonal defect. This topological obstruction is resolved by systematically applying right rotations $\mathcal{R}$ to subsequent segments, which restores sublattice registry. The complete circumferential path of the nanotube, incorporating all six pentagons, is therefore:
\begin{equation}
\overset{\leftrightarrow}{\mathbf{P}}_{\text{CNT}} = \sum_{t=1}^{6} \mathcal{R}^{t-1} (\mathbf{P}_{t \to t+1} + \mathbf{P}_5),
\label{eq:1}
\end{equation}	
where $\mathcal{R}^{t-1}$ denotes $t-1$ successive applications of the $\mathcal{R}$ operator.

Given that three right rotations invert any path, $\mathcal{R}^3(\mathbf{P}) = \mathbf{P}^{-1}$, the pentagon contributions sum to identity: $\sum_{t=1}^{6} \mathcal{R}^{t-1}(\mathbf{P}_5) = \mathbf{e}_0$. Substituting the explicit path forms yields:
\begin{equation}
\overset{\leftrightarrow}{\mathbf{P}}_{\text{CNT}} = \sum_{t=1}^{6} \mathcal{R}^{t-1} (n_{t} \mathbf{p}_{ji} + m_{t} \mathbf{p}_{jk}).
\label{eq:2}
\end{equation}	
To simplify the expression in Equation (2), we exploit the identity $\mathbf{P} = \mathcal{R}^{t-1}[\mathcal{L}^{t-1}(\mathbf{P})]$ and re-parameterize each $(n_t, m_t)$ in successively left-rotated topological spaces. This aligns all pentagon contributions into a common reference frame:
\begin{equation}
n_t \mathbf{p}_{ji} + m_t \mathbf{p}_{jk} = \dot{n}_t \mathcal{L}^{t-1}(\mathbf{p}_{ji}) + \dot{m}_t \mathcal{L}^{t-1}(\mathbf{p}_{jk}).
\label{eq:3}
\end{equation}
The closed nanotube path defines its chirality:
\begin{equation}
\overset{\leftrightarrow}{\mathbf{P}}_{\text{CNT}} = n \cdot \mathbf{p}_{ji} + m \cdot \mathbf{p}_{jk}.
\label{eq:4}
\end{equation}		
Substituting Equations (\ref{eq:3}) and (\ref{eq:4}) into Equation (\ref{eq:2}) gives:
\begin{equation}
n \cdot \mathbf{p}_{ji} + m \cdot \mathbf{p}_{jk} = \left( \sum_{t=1}^{6} \dot{n}_t \right) \mathbf{p}_{ji} + \left( \sum_{t=1}^{6} \dot{m}_t \right) \mathbf{p}_{jk}.
\label{eq:5}
\end{equation}

\begin{figure*}[htb]
       \begin{center}
              \includegraphics[width=\textwidth]{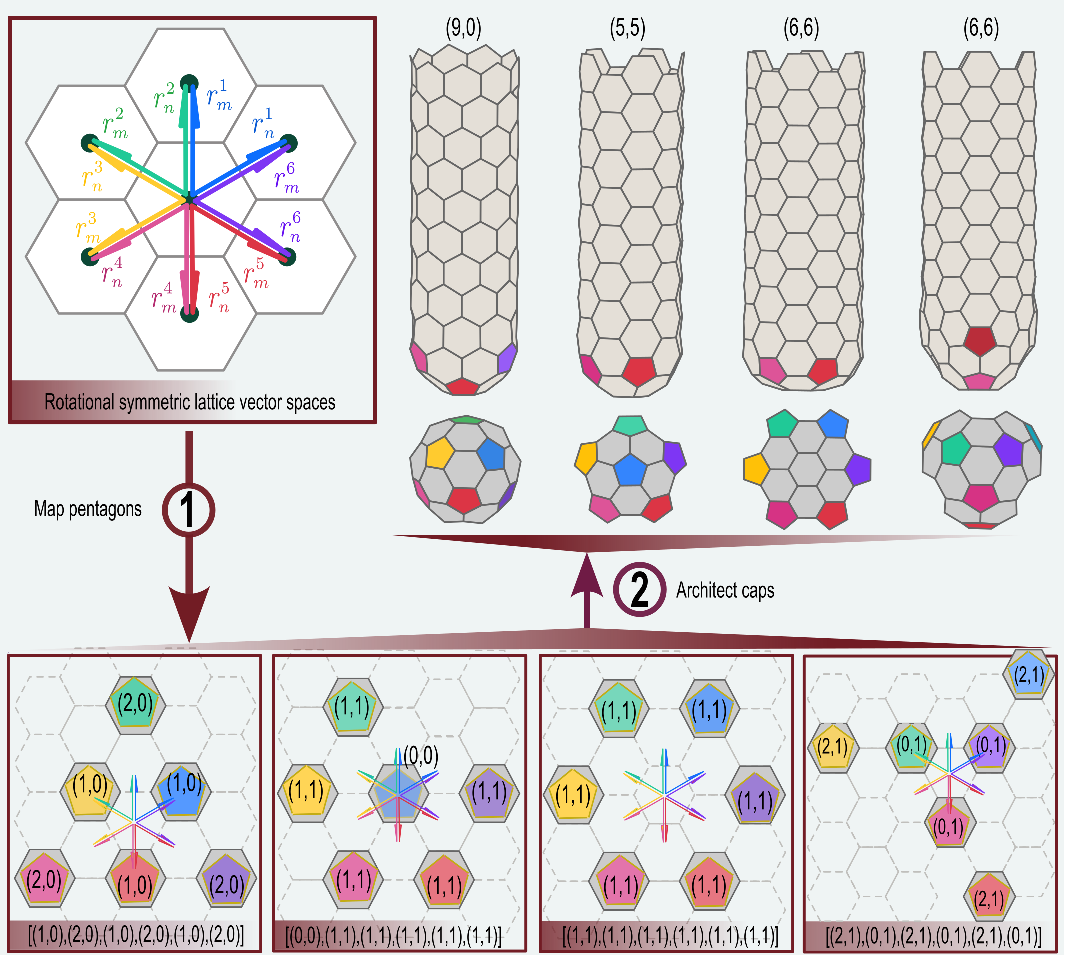}
       \end{center}
       \caption{Chiral index assignment of a carbon nanotube cap. (Top left) Six rotationally generated vector spaces obtained by successive left rotations. (Bottom) Chiral indices of the cap mapped onto the graphene lattice in corresponding vector spaces. (Top right) Atomic configuration of the cap and the resulting carbon nanotubes.}
       \label{fig:2}
\end{figure*}

Equation (\ref{eq:5}) reveals the fundamental linear relationship. The Abelian group $V$ is isomorphic to the group generated by physical lattice vectors $\mathbf{r}_n$, $\mathbf{r}_m$. Here, $\mathbf{p}_{ji}$ and $\mathbf{p}_{jk}$ are their topological proxies; a left rotation $\mathcal{L}$ corresponds to a $\pi/3$ real-space rotation of the lattice basis. We therefore introduce a chiral index notation for the cap: $[(\dot{n}_1, \dot{m}_1), (\dot{n}_2, \dot{m}_2), \ldots, (\dot{n}_6, \dot{m}_6)]$, where each tuple $(\dot{n}_t, \dot{m}_t)$ represents the coordinates of a pentagon within the lattice space $(\mathbf{r}_{n}^{t},\mathbf{r}_{m}^{t})$ (Fig.~\ref{fig:2}). Treating these notation tuples as integer vectors leads to the final vector sum rule:
\begin{equation}
(n, m) =  \sum_{t=1}^{6} \left(\dot{n}_t, \dot{m}_t \right).
\label{eq:6}
\end{equation}

Equation (\ref{eq:6}) presents a central result of this work: the vector sum rule governing carbon nanotube chirality. This rule dictates that the chiral indices $(n, m)$ are determined entirely by the vector sum of the coordinates of the six pentagons, evaluated across rotationally related lattice vector spaces. A profound implication is that while a given cap can be represented by multiple notation sequences due to the symmetry of the hexagonal lattice, the resulting sum in Eq.~(\ref{eq:6}) remains invariant under all symmetry operations of the P6/mmm space group. This means that although the specific notation sequences of a cap may be altered by symmetry operations—including translation, rotation, and reflection—the topological sum rule itself remains unchanged. This invariance underscores the rule's fundamental character as a universal topological law. 

The vector sum rule provides not only an intuitive bridge between cap structure and nanotube chirality but also profound physical insight into its origin. We propose a topological interpretation in which the graphene edge is taken as a reference state, since any closed path on its lattice is homotopic to the identity element $\mathbf{e}_0$. Each pentagon, however, acts as a topological deflector, rotating the growing edge path by $\pi/3$---a direct consequence of the sublattice mismatch introduced by the pentagonal defect. After six discrete rotations, the crystalline registry is restored but the edge is redirected to form a nanotube circumference. The chirality of the resulting nanotube is determined by the spatial distribution of these six discrete rotations within the hexagonal lattice.

Due to the translational symmetry of the hexagonal lattice, fully defining a cap requires exactly 10 integers when one pentagon is taken as a reference position. Interestingly, a cap contains significantly more structural information than its resulting nanotube edge, which is completely described by only two integers $(n,m)$. As a consequence, each chirality corresponds to a large number of distinct cap configurations. For example, $(12,6)$ nanotubes can theoretically nucleate from 22,127 different caps~\cite{Brinkmann1999}, revealing substantial structural degeneracy in the nucleation process.

\section*{Deterministic Nucleation}

Two distinct pathways—clone growth and solid catalyst–specified growth via CVD—offer routes toward chirality control over nanotube synthesis. Clone growth directly programs the cap architecture through molecular-scale design. For instance, chemical opening of $\mathrm{C}_{60}$ molecules yields well-defined cap sequences such as $[(1,0),(2,0),\ldots]$ and $[(0,0),(1,1),\ldots]$, which template the growth of $(9,0)$ and $(5,5)$ nanotubes, respectively (Fig.~\ref{fig:2})~\cite{Yu2010}. Alternatively, molecular templates synthesized through organic chemistry—such as the $\mathrm{C}_{96}\mathrm{H}_{54}$ molecule, which preorganizes the sequence $[(2,1),(0,1),\ldots]$—can direct cap assembly to selectively form $(6,6)$ nanotubes (Fig.~\ref{fig:2})~\cite{Sanchez-Valencia2014}. Clone growth provides a clear demonstration of the direct pathway from nucleation to nanotube formation. However, it suffers from slow elongation rates beyond the microscale. 

In contrast, CVD using solid catalysts enables rapid elongation, but nucleation remains poorly understood, despite having produced several nanotubes with high enantiomeric purity~\cite{Yang2022b,He2019}. Among these high-purity species, the $(12,6)$ species exhibits notable and reproducible abundance across a wide range of synthesis conditions~\cite{Che2014,Yang2014,Zhang2016}. Although kinetic theory accounts for its rapid elongation, the preferential nucleation of $(12,6)$ nanotubes has remained an experimental puzzle. Here, we address this long-standing question through a topological framework.

We begin by examining a fundamental dichotomy in chirality-controlled nanotube synthesis: is chirality statistically determined by post-nucleation kinetic processes, or deterministically encoded through selective cap formation? While subtle, this distinction carries profound implications. The prevailing assumption of former has directed extensive attention toward post-nucleation chirality control via edge-matching mechanisms. This perspective, however, faces a critical challenge: each chirality corresponds to numerous distinct cap configurations, a number that grows rapidly with diameter, as does the diversity of possible chiralities. Indeed, the configurational entropy of possible cap structures is effectively infinite. Consequently, any uncontrolled nucleation process would yield the complete spectrum of cap structures and, therefore, all possible nanotube chiralities. Even with diameter constraints, numerous chiralities exhibiting similar favorable post-nucleation growth conditions would be nucleated at same time. Thus, post-nucleation mechanisms alone cannot adequately explain the observed selective chirality enrichment. The origin of chirality control must therefore originate from deterministic nucleation processes where only specific cap structures are selectively formed. The solid catalyst serves as a template that dramatically narrows the vast configuration space to a limited set of specific structures—as unequivocally demonstrated by the precise structural templating observed in clone growth experiments \cite{Yu2010,Sanchez-Valencia2014}.The challenge then becomes identifying the specific cap architectures responsible for the preferential formation of $(12,6)$ nanotubes and understanding the underlying physical principles that favor such configurations. The vector sum rule provides the foundation to systematically enumerate all possible caps satisfying this chirality condition under catalyst constraints, enabling targeted investigation of their relative stabilities and catalytic matching properties.

\begin{figure*}[htb]
              \begin{center}
                     \includegraphics[width=1.0\linewidth]{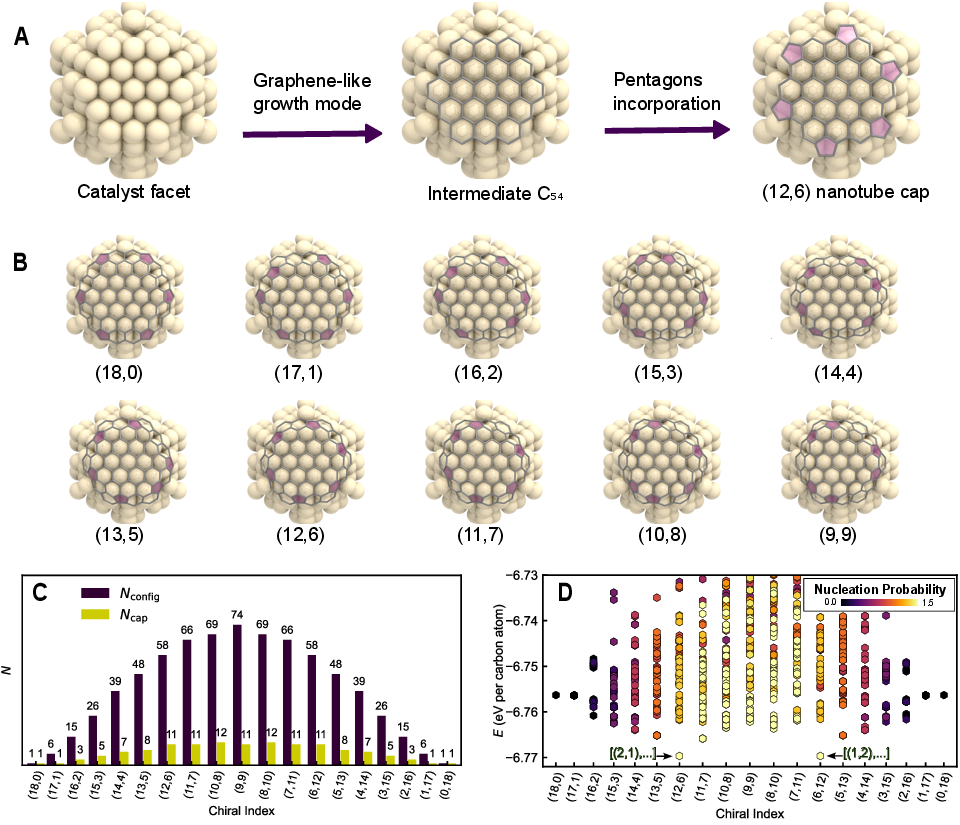}
              \end{center}
              \caption{A. Schematic illustrations of cap formation on $\langle 1,1,1 \rangle$ crystallographic facets of catalyst nanoparticles. B. Representative cap structures corresponding to distinct chiralities derived from the $\mathrm{C}_{54}$ intermediate. C. Distribution of $\mathrm{C}_{54}$-derived caps mapped in chirality space. (D) Per-carbon-atom formation energy of caps on a copper catalyst as a function of chiral angle $\theta$, with color scale representing nucleation probabilities.}
              \label{fig:3}
\end{figure*}

The enrichment of $(12,6)$ nanotubes is predominantly observed on $\langle 1,1,1 \rangle$ facets of face-centered-cubic catalysts (e.g., Cu, Co, Ni, and Pt nanoparticles)~\cite{Yang2020}. These catalysts possess a lattice parameter of approximately 2.5 $\AA$, closely matching graphene's 2.47 $\AA$ lattice constant---making these six-fold symmetric facet ideal for templating carbon nanostructures. Under such geometric constraints, the $[(2,1),\ldots]$ cap structure emerges due to its inherent symmetry matching, establishing a direct correlation with $(12,6)$ nanotube enrichment. The $\langle 1,1,1 \rangle$ facets exert precise control over initial cap formation, selectively directing nucleation toward $[(2,1),\ldots]$ configurations (Fig.~\ref{fig:3}A). During this process, the carbon network extends epitaxially across the catalyst facet in a graphene-like growth mode, forming a $\mathrm{C}_{54}$ intermediate composed of 19 hexagonal carbon rings. This structural preference is consistently supported by graphene growth studies~\cite{Artyukhov2015,Zhang2024}. The critical transition to cap formation occurs when the expanding carbon network encounters the catalyst facet boundary, triggering pentagon incorporation that yields the specific $[(2,1),\ldots]$ cap---a well-established mechanistic step~\cite{Huang2011,Pigos2011,Yang2020,Swinkels2023}.

The enrichment of $(12,6)$ nanotubes can thus be explained by the combination of favorable nucleation characteristics with optimal post-nucleation growth kinetics—specifically, a diameter of approximately 1.3~nm and a chiral angle of $19.1^\circ$~\cite{Ding2008,Qiu2021,He2019b}.

However, achieving such well-defined growth requires precise incorporation of all six pentagons at $(2,1)$ sublattice sites (or equivalently, $(1,2)$ sites, which yield the mirror-symmetric $(6,12)$ chirality). Any deviation from this ideal configuration direct the grow to alternative chirality. For example, substituting two $(2,1)$ pentagons with $(3,0)$ sites—deterministically shifts the resulting chirality to the semiconducting $(14,4)$ type. Consistent with this theoretical framework, the $(14,4)$ nanotube exhibits experimental enrichment under similar synthesis conditions~\cite{Yang2016}. Moreover, several other structurally related chiralities—including $(18,0)$, $(16,2)$, $(15,3)$, and $(9,9)$—are also enriched under specific synthesis conditions(Fig.~\ref{fig:3}B)~\cite{Yang2020}.

A fundamental question persists: why do $(12,6)$ nanotubes exhibit such overwhelming dominance under typical synthesis conditions? Consider a solid catalyst with a diameter of 1.4~nm---a common size in nanotube growth---whose surface possesses a facet of 19 atoms arranged in a honeycomb lattice. This specific facet template dictates the exclusive formation of the $\mathrm{C}_{54}$ carbon cluster. The periphery of this cluster presents 18 active sites for pentagon incorporation, generating 730 possible cap configurations ($N_\text{config}$) that satisfy the isolated pentagon rule (Fig.~\ref{fig:3}C). These correspond to 70 structurally distinct caps  ($N_\text{cap}$) , which in turn seed 10 different nanotube chiralities. If pentagon incorporation were a random process, the resulting chirality distribution would mirror the statistical weights of these configurations---a prediction starkly at odds with experimental evidence. The pronounced enrichment of $(12,6)$ nanotubes from this vast configurational space unequivocally demonstrates that decisive energetic and kinetic factors override entropic considerations.

As revealed by \textit{in situ} transmission electron microscopy~\cite{Zhang2024}, nanotube edges undergo repeated etching and regrowth during nucleation, progressively evolving toward lower energy states. At this stage, low-energy cap configurations gain a significant advantage over alternatives. As shown in Fig.~\ref{fig:3}C, the $[(2,1),\ldots]$ cap exhibits the lowest formation energy among all 730 possible configurations. This distinctive advantage stems from two key factors: maximized inter‑pentagon distances that reduce strain energy, and improved registry with the catalyst surface that minimizes interfacial energy. 

However, whether this nucleation process reaches thermodynamic equilibrium remains heavily debated. Previous studies have indeed identified kinetic barriers in the growth of some chiral tubes~\cite{Artyukhov2014}. Consequently, confirming the absence of kinetic barriers in the nucleation pathway for the $[(2,1),\ldots]$ cap is essential. Experimental evidence indicates that the addition of a $\mathrm{C}_3$ sequence to an active site is significantly less favorable than the incorporation of a $\mathrm{C}_2$ unit. Notably, from a purely kinetic standpoint, pentagon incorporation is generally favored over hexagon formation in most scenarios~\cite{Swinkels2023}.

The first application of kinetic theory to nanotube growth rate analysis was pioneered by Yakobson et al.~\cite{Ding2008}, who modeled growth rates as proportional to the number of active sites available for $\mathrm{C}_{2}$ addition. We extend this framework by defining the probability of a $\mathrm{C}_{2}$ addition event as $\phi(t)=1-\mathrm{e}^{-\lambda t}$, where $\lambda$ is an experimentally determined rate constant and $t$ is time. Cap formation is considered complete when a new edge is generated on the existing $\mathrm{C}_{54}$ intermediate. To simplify the model, we consider only $\mathrm{C}_{2}$ additions at active sites and treat $\lambda$ as constant for all events. A fundamental constraint is that hexagon incorporation becomes feasible only when an adjacent site has already incorporated a pentagon or hexagon. This leads to an Erlang probability density for completing an $N$-subsequence incorporation at time $t$, described by $\Theta(t,N)=\lambda^{N} t^{N-1} \mathrm{e}^{-\lambda t} / \Gamma (N)$. Each local incorporation event culminates in a hexagon at the $(3,0)$ position, requiring pre-incorporation of neighboring sublattices from both directions. The probability density for this sub-event at time $t$ is given by
\[
\Psi(t,\alpha,\beta) = \int_{0}^{t} \Theta(\tau,\alpha) \Theta(\tau,\beta) \lambda \mathrm{e}^{-\lambda (t-\tau)}  d\tau,
\]
where $\alpha$ and $\beta$ represent the number of incorporations from each respective direction.

\sloppy
In the simplest case of the $[(3,0),(3,0),(3,0),(3,0),(3,0),(2,1)]$ cap—corresponding to the $(17,1)$ tube—incorporation initiates at the $(2,1)$ sublattice pentagon site and propagates sequentially around the circumference until reaching the final $(3,0)$ sublattice. The cumulative probability distribution is given by $F_{(17,1)}(t) = \int_{0}^{t} \Psi(\tau,16,1) \, d\tau$.  

For the symmetric $[(2,1),\ldots]$ cap associated with the $(12,6)$ tube, edge expansion begins independently at all six $(2,1)$ sites, culminating in hexagon formation at each $(3,0)$ position. The cumulative formation probability by time $t$ follows as  
\[
F_{(12,6)}(t) = \left[\int_{0}^{t} \Psi(\tau,2,1) \, d\tau\right]^{6},
\]  
where the integral reflects the cumulative probability for each individual growth arm.

In the special case of the $[(3,0),\ldots]$ cap corresponding to the $(18,0)$ tube, a $\mathrm{C}_{3}$ addition is required to form the initial hexagon or pentagon, a necessary step for propagation. We therefore treat its incorporation probability as effectively negligible.

Our kinetic model reveals a pronounced advantage of the $[(2,1),\ldots]$ cap structure over most alternative configurations (Fig.~\ref{fig:3}D, parameters: $\lambda=10^{3}$, $t=10^{-2}$, corresponding to a typical nucleation timescale of 10~ms). As nanotube growth is governed by both thermodynamic stability and kinetic accessibility, we demonstrate conclusively that the dominance of the $[(2,1),\ldots]$ cap drives the preferential formation of $(12,6)$ nanotubes.

Within this framework, a minority of cap structures are predicted to have marginally higher formation probabilities than the symmetric $[(2,1),\ldots]$ cap. A representative case is the $[(2,1),(1,2),\ldots]$ cap of the armchair $(9,9)$ tube. This apparent overestimation stems from pathways involving hexagon incorporation between neighboring pentagons---a event that necessitates a substantial structural reorganization and is thus expected to be energetically costly, resulting in a reduced $\lambda$ value. A predictive kinetic model must therefore integrate the complete ensemble of structural variations while accounting for the influence of dynamic catalyst morphology on incorporation probabilities. The pivotal role of catalyst morphology has been extensively demonstrated in the synthesis of pristine graphene through \textit{in situ} transmission electron microscopy studies~\cite{Huang2011}. While pentagon-heptagon pairs represent the dominant defects in graphene, nanotube nucleation is governed exclusively by pentagon formation~\cite{Reich2006,Sun2025}. The dynamics of pentagon incorporation during nucleation remain largely unexplored and warrant detailed investigation.


\section*{Conclusion}
		
Through the discovery of the vector sum rule and its application to six-fold symmetric nucleation caps, we have reframed chirality-controlled nanotube synthesis as a deterministic process governed by cap topology and epitaxial matching with catalyst surfaces. While previous theories attributed chirality-selective growth primarily to edge matching mechanisms~\cite{Magnin2018}, our findings establish that nucleation is deterministically controlled by topological and interfacial factors. This framework coherently explains the experimental enrichment of $(12,6)$ nanotubes while accounting for chiral enrichment in neighboring species. The parallel enrichment observed around $(6,6)$ nanotubes—which similarly feature six-fold symmetric caps—further validates this mechanism~\cite{Chu2023}. Although our work captures essential nucleation features, developing predictive synthesis requires a more comprehensive understanding of how catalyst morphology directs pentagon formation. Our results provide the theoretical foundation for advancing toward this goal.

\section*{Acknowledgement}

This study was supported by the National Natural Science Foundation of China (Nos. 12462009 and 12132008). We thank Nader Anani for advice on the manuscript. 

\subsection*{Data, Materials, and Software Availability}
\sloppy
All data are available in the main text or the supplementary materials. A direct implementation of the topological framework into computer code for designing carbon caps is available at GitHub (https://github.com/ZhengrongGuoChina/Graphene-topology).

\subsection*{Method}

The formation energies in Fig.~\ref{fig:3}D were calculated using the LAMMPS package. Carbon-carbon interactions were described by the reactive bond-order (REBO) potential, while carbon-copper substrate interactions were modeled using a Lennard-Jones potential ($\sigma = 2.89$~\AA, $\epsilon = 16.2$~meV). All simulations were performed in the canonical (NVT) ensemble at 1~K with a rigid substrate 

\label{Bibliography}
\bibliographystyle{unsrtnat}
\bibliography{citations}

\end{multicols}
\end{document}